\documentclass[aps,preprint,showpacs,showkeys]{revtex4}
\usepackage{graphicx,epsfig}
\usepackage{amssymb}

\begin{document}

\title{Effects of quintessence on scattering and absorption sections of black holes}

\author{L. A. L\'opez } 
\email[Corresponding author, E-mail:] {lalopez@uaeh.edu.mx}
\author{Omar Pedraza}

\affiliation{$^1$\'Area Acad\'emica de Matem\'aticas y F\'isica, UAEH, Carretera Pachuca-Tulancingo Km. 4.5, C P. 42184, Mineral de la Reforma, Hidalgo, M\'exico.}

\begin{abstract}
Based on the ideas used by Kiselev, we study three black holes surrounded by quintessence and the effects of quintessence on the classical and semiclassical scattering cross-sections. In contrast, the absorption section is studied with the sinc approximation in the eikonal limit. For Schwarzschild, Reissner-Nordstr\"{o}m and Bardeen black holes surrounded by quintessence, the critical values of charges and the normalization factor are obtained. We also described the horizons and the extremal condition of the black holes surrounded by quintessence, by setting the quintessence state parameter in the two particular cases $\omega=-2/3$ and $\omega=-1/2$.
\end{abstract}
\maketitle

\section{Introduction}\label{1}

Black holes are objects that can be described by their mass, angular momentum, and charge. Black holes are the subject of many studies because the observational data currently available suggest that said objects are distributed in all the galaxies in the Universe \citep{Nar05}, and they provoke fascinating theoretical questions such as the Information Paradox.

For the study of the physical properties of black holes, the test fields around them are analyzed. For example, an important aspect is the study of geodesics \citep{Pug17} around black holes with different configurations. Also, the quasi-normal modes resulting when a black hole (BH) undergoes perturbations \citep{Kon11} are analyzed. The absorption and scattering of matter and fields has an important role in the phenomenology of BHs, and has been studied in many configurations of BHs \citep{Dol08,Mac16}.

In the classical scattering studies one can consider parallel null geodesics coming from infinity with a critical impact parameter \citep{Col73}. The semi-classical scattering is analyzed when we consider the interference that occurs between scattered partial waves with different angular momenta (Glory approximation \citep{Mat85}). In the sinc approximation, the absorption cross-section is proportional to the sum of the called geometric cross-section and the oscillatory part of the absorption cross-section \citep{Dec11}.

Since black holes are objects found throughout the universe, they also can interact with different types of matters and fields. Considering the dark energy dominating the universe, the BHs are surrounded by dark energy. 

There are different models as candidates for dark energy. Most of them are based on a scalar field. The difference between the models is the magnitude of the state parameter $\omega$ which is the ratio of pressure to energy density of dark energy. For quintessence $-1<\omega<-1/3$, and when we consider the cosmological constant $\omega=-1$.

In order to study the black holes surrounded by dark energy, Kiselev (2003) \citep{Kis03} presented new static spherically symmetric exact solutions of the Einstein equations for black holes surrounded by quintessence. The solution of Schwarzschild surrounded by quintessence is studied in \citep{Fer12} where the null geodesics are analyzed. The thermodynamics of the Schwarzschild and Reissner-Nordstr\"{o}m BHs surrounded by quintessence is discussed in \citep{Gha16}. Also, the thermodynamics and null geodesics of a Bardeen BH surrounded by quintessence in \citep{Ghaderi:2017yfr} are studied. In \citep{Ghosh:2015ovj} is derived a rotational analogue of  Schwarzschild BH surrounded by quintessence by employing the Newman–Janis complex transformation.

Other types of black holes have been analyzed when the quintessence is considered (for example see \citep{Malakolkalami:2015cza,Zheng:2019mem}), the studies have focused on thermodynamic properties \citep{Rodrigue:2018lzp,Liang:2020uul}, or quasi-normal modes \citep{Saleh:2018hba} and geodesic properties. All of them have shown modifications of their proposals, so it is clear that the scattering and absorption cross-sections will also be modified when considering quintessence. 

Motivated by the investigations about the effects of quintessence in black holes and the comparison between them  (for example, see \citep{Ghaderi:2017wvl}), we studied how the effects of quintessence modify the classical and semi-classical scattering sections as well as the absorption section in the sinc approximation. 

The paper is organized as follows. In Sec. \ref{2}, using the Kiselev ideas, three black holes surrounded by quintessence are presented, also the charges critical and normalization factors critical are shows and we  analyzed  the event horizons and the cases extremal. In \ref{3}, the expressions for the classical and semi-classical scattering sections are presented and we analized the sections for the threes black holes surrounded by quintessence consider two values of the normalization factor quintessence state parameter. Section \ref{4}. taking the same values the adsorption section is obtained by sinc approximation. Finally in \ref{5}, the conclusions are given.

\section{Black holes surrounded by quintessence}\label{2}

In 2003 Kiselev \citep{Kis03}  proposed new static spherically symmetric solutions that describe Black holes surrounded by quintessence, considering that the energy-momentum tensor for quintessence should satisfy \citep{Cvetic:2016bxi};

\begin{equation}
T_{\phi}^{ \phi}=T_{\theta}^{ \theta}=-\frac{1}{2}(3\omega +1)T_{r}^{ r}=\frac{1}{2}(3\omega +1)T_{t}^{t},
\end{equation}

where $\omega$ is taken to be a constant and  the dominant energy condition requires $\rho=T_{tt}\geq 0$, when $\rho$ is the energy density with equation of state $p=\rho \omega $ ($p$ is the pressure) and $\mid 3\omega + 1 \mid \leq 2$.

The energy density is given by;

\begin{equation}
\rho=-\frac{c}{2}\frac{3\omega}{r^{3(1+\omega)}}
\end{equation}

Where $c$ is a normalization factor, the quintessence pressure must be negative to accelerate the universe. Since the energy density should be positive, the normalization factor must be positive for a negative $\omega$.

The line element that considers solutions surrounded by quintessence is given by;

\begin{equation}\label{mfa}
ds^2=-f(r)dt^2+f^{-1}(r)dr^2+r^2d\theta^2+r^{2}\sin^2\theta d\phi^{2},
\end{equation}
where
\begin{equation}
f(r)=1-\frac{M(r)}{r^{2}}-\frac{c}{r^{3\omega+1}}.
\end{equation}

Here, $M(r)$ is a function that contains the parameter of mass and other parameters (as electric or magnetic charge), and $\omega$ is within the range $-1<\omega<-1/3$ and represents the quintessence parameter.

Kretschmann scalar $K$ for the solution (\ref{mfa}) is given by;

\begin{equation}
K=R_{\mu\nu\rho\sigma}R^{\mu\nu\rho\sigma}= \frac{A(r)}{r^{(6\omega+6)}},
\end{equation} 

where $A(r)$ is a polynomial function of $r$ with  $A(r = 0) \neq  0$. Therefore it can be concluded that a physical singularity occurs at $r = 0$ if $\omega \neq \{-1\}$.

For the Schwarzschild (Schw), Reissner-Nordstr\"{o}m (RN) and Bardeen black holes surrounded  by quintessence (BHs-$\omega$), $f(r)$ is given by:

\begin{equation}\label{fs}
f(r)_{Schw}=1-\frac{2M}{r}-\frac{c}{r^{3\omega+1}},
\end{equation}
\begin{equation}\label{fr}
f(r)_{RN}=1-\frac{2M}{r}+\frac{q^{2}}{r^{2}}-\frac{c}{r^{3\omega+1}},
\end{equation}
and
\begin{equation}\label{fb}
f(r)_{Bar}=1-\frac{2M}{r}+\frac{3Mg^{2}}{r^{3}}+O(r^{5})-\frac{c}{r^{3\omega+1}}.
\end{equation}

Where $M$ is the parameter of mass, $q$ the electric charge, and $g$ the magnetic charge of a nonlinear self-gravitating monopole, in the case of the Bardeen black hole. In \citep{Gha16} the authors studied the thermodynamic quantities for Bardeen BH-$\omega$ assuming that the terms of higher orders of $O(r^{5})$ can be neglected and give the same behavior, given that if we plot the asymptotical form of the Bardeen black hole,  we can see that it has the same number of horizons and it is asymptotically flat. 

On the other hand, for the parameter $ \omega=-1$, the metrics describe de-Sitter black holes.

The Schwarzschild BH surrounded by quintessence is studied in \citep{Fer12} for the particular case of $\omega=-2/3$. The Null geodesics of RN BH with quintessence are boarded in \citep{Malakolkalami:2015tsa}. And in \citep{Dariescu:2021jjh}, the timelike geodesics of RN BH-$\omega$ considering $\omega=-2/3$ are studied.   

The horizons are determined by the positive roots of the equation $f(r)=0$. For this analysis, we express radial distance and the parameter $c$ in units of mass as; $c\to c/M^b$ $r\to r/M$. In the case of the charges, the RN BH-$\omega$ electric charge goes as  $q\to q/M$, and for Bardeen BH-$\omega$ the magnetic charge goes as $g\to g/M$, with $b=3\omega+1$, then the event horizons are the roots of:

\begin{equation}\label{fs1}
(cr+2r^{b}-r^{1+b})r^{-(b+1)}=0_{Schw},
\end{equation}
 
\begin{equation}\label{fR1}
(cr^{2}-q^{2}r^{b}+2r^{1+b}-r^{b+2})r^{-(b+2)}=0_{RN},
\end{equation}

\begin{equation}\label{fb1}
(cr^{3}-3g^{2}r^{b}+2r^{b+2}-r^{b+3})r^{-(b+3)}=0_{Bar}.
\end{equation}

The number of horizons depends entirely on the choice of the values of parameters $\omega$, $c$, $g$ and $q$, when considering the quintessence term, a new horizon emerges, the cosmological (quintessence) horizon (for example see \citep{Rizwan:2018lht,Ghaderi:2017wvl}). If the RN BH and Bardeen BH have two horizons, then when we consider the parameter of quintessence, the BHs-$\omega$ could have three horizons. In the case of Schwarzschild, with quintessence we can have two horizons.

Now we determine the range of values for $q$, $g$ and $c$, so that the different line elements represent a black hole or an extremal black hole, for this, we make use of  the method described in \citep{Toshmatov:2015npp,Pedraza:2020uuy}. First, when we consider $f(r_{h})=0$ (Eqs. (\ref{fs1}), (\ref{fR1}) and  (\ref{fb1})) the $q$ and $g$ parameters can be parametrized as a function of $r_{h}$ (horizon) and $c$.
 
\begin{equation}\label{QR}
q^{2}(r_{h},c)=(cr_{h}+2r_{h}^{b}-r_{h}^{1+b})r_{h}^{1-b},
\end{equation}

\begin{equation}\label{Gb}
g^{2}(r_{h},c)=\frac{r_{h}^{2-b}}{3}(cr_{h}+2r_{h}^{b}+r_{h}^{b+1}).
\end{equation}

In the case of Schwarzschild  BH-$\omega$ (Eq. (\ref{fs1})), it has an extrema value for $c(r_{h})_{Schw}$, the Eqs. (\ref{QR}) and (\ref{Gb}) take extremal values in $c(r_{h})_{RN}$ and $c(r_{h})_{Bar}$ given by;

\begin{eqnarray}\label{C1}
c(r_{h})_{Schw}&=&(r_{h}-2)r_{h}^{b-1}, \nonumber\\
c(r_{h})_{RN}&=&\frac{2(r_{h}-1)r_{h}^{b-1}}{2-b},\nonumber\\
c(r_{h})_{Bar}&=&\frac{(3r_{h}-4)r_{h}^{b-1}}{3-b}.
\end{eqnarray}
 
Now, the different $c(r_{h})$ (\ref{C1}) have extremes $\left(\frac{dc(r_{h})}{dr_{h}}=0\right)$ in a certain $r_{crit}$, performing an analysis we obtain; 

\begin{eqnarray}\label{rcr}
r_{crit_{Schw}}&=&\frac{2(b-1)}{b} ,\nonumber\\
r_{crit_{RN}}&=&\frac{b-1}{b}, \nonumber\\ 
r_{crit_{Bar}}&=&\frac{4(b-1)}{3b}.
\end{eqnarray}

Since the parameter $c>0$, $c(r_{crit})$ must be positive in the range $b\in(-2,0)$.

Then the critical values of the different parameters $c(r_{crit})$ are given in terms of $\omega$ as; 

\begin{equation}
c(r_{crit})_{Schw}=c_{crit_{Schw}}=-\left( \frac{6\omega}{1+3\omega}\right)^{3\omega}\left( \frac{2}{1+3\omega} \right),
\end{equation}

\begin{equation}\label{Cw}
c(r_{crit})_{RN}=c_{crit_{RN}}=27^{\omega}\left( \frac{\omega}{1+3\omega}\right)^{3\omega}\frac{2}{9\omega^{2}-1} ,
\end{equation}

\begin{equation}\label{Cw1}
c(r_{crit})_{Bar}=c_{crit_{Bar}}=\left(\frac{4\omega}{1+3\omega}\right)^{1+3\omega}\frac{1}{\omega(3\omega-2)},
\end{equation}

the charge parameters $q^{2}_{c}=q^{2}(c_{crit_{RN}},r_{crit_{RN}})$ and $g^{2}_{c}=g^{2}(c_{{crit_{Bar}}},r_{crit_{Bar}})$ are given by;

\begin{equation}\label{Mw}
 q^{2}_{c}=\frac{9\omega^{2}}{9\omega^{2}-1}, \;\;\;\;\ g^{2}_{c}=\frac{32\omega^{3}}{3(3\omega-2)(3\omega+1)^{2}}.
\end{equation}

In summary, for $0 \leqslant c \leqslant c_{crit}$, $0 \leqslant g^{2} \leqslant g^{2}_{c}$  and $0 \leqslant q^{2} \leqslant q^{2}_{c}$ ,the different BHs-$\omega$ can represent a black hole with different horizons $r_{in}$, $r_{out}$ and $r_{\omega}$ (quintessence horizon), in the case of Schw BH-$\omega$, it has two horizons or one horizon.

\begin{figure}[h]
\begin{center}
\includegraphics [width =0.45 \textwidth ]{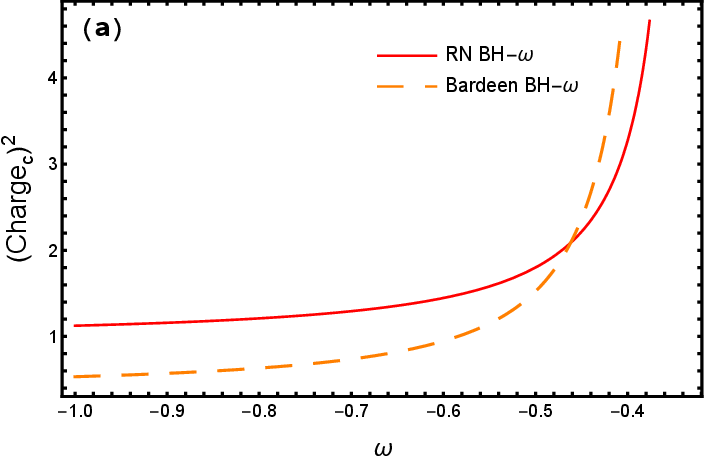}
\includegraphics [width =0.45 \textwidth ]{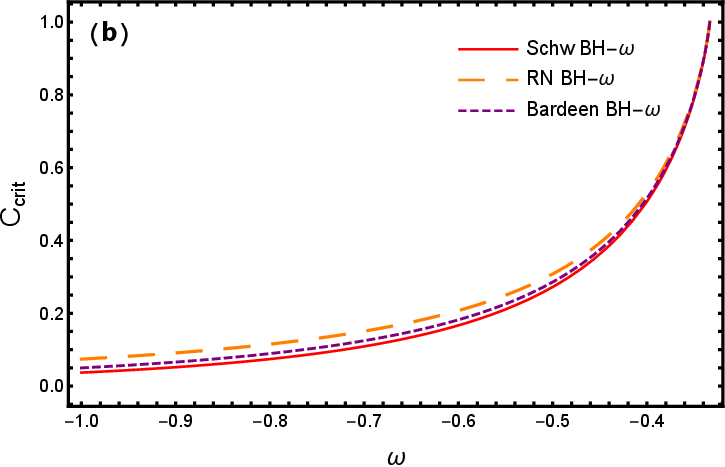}
\end{center}
\caption{{\bf (a)} Behavior of $q^{2}_{c}$, $g^{2}_{c}$ as a function of $\omega$. {\bf (b)} Behavior of $c_{crit}$ as a function of $\omega$.}\label{Fig1}
\end{figure}

The behavior of the critical values $q^{2}_{c}$  and $g^{2}_{c}$ versus $\omega$ for the different BHs-$\omega$ is shown in Fig. \ref{Fig1} {\bf (a)}. In the Fig. \ref{Fig1} {\bf (b)} the behavior of $c_{crit}$ is shown.

The Fig. \ref{Fig1} {\bf (b)} shows that as $\omega$ increases $c_{crit}$ increases for the different BHs-$\omega$, when $\omega \rightarrow -1$, the values of $c_{crit}$ are finite. On the other hand,when $\omega \rightarrow -1/3$, we obtain that $q_{c}^{2}\rightarrow \infty$ and $g_{c}^{2}\rightarrow \infty$ (see Fig. \ref{Fig1} {\bf (a)}), this means that in the presence of a quintessence field, a higher charge black hole is formed.

It is also possible to discuss the behaviour of extremal black holes, and naked singularities, represented by the line elements  (\ref{fs}), (\ref{fr}) and  (\ref{fb}). Since when only one horizon remains,  (\ref{fs}), (\ref{fr}) and  (\ref{fb}) describe a naked singularity.

We propose to choose the value of $\omega =-2/3$, which provides an intermediate range of  $c$, $q_{c}$ and $g_{c}$. An interesting point that Kiselev noted in the case $\omega=-2/3$, is that the BHs-$\omega$ are not asymptotically flat but have an asymptotic defect angle.  Also, the case $\omega =-1/2$  enable a relatively simple treatment of the properties of BHs-$\omega$. 

\subsection{Black holes with $\omega=-2/3$}
 
We will focus on the particular case of $\omega= -2/3$. The Schw BH-$\omega$ has two horizons for $8c<1$ (also see \citep{Fer12}), but the cases of RN BH-$\omega$ and Bardeen BH-$\omega$ can have three or two horizons. As our intention is to compare the different black holes, then we approach to the extremal condition. 

The RN BH-$\omega$ extremal can be obtained when the conditions $f(r)_{RN}=0$ and $\frac{d}{dr}f(r)_{RN}=0$ are satisfied simultaneously. Introducing $q^{2}(r,c)_{RN}$ of (\ref{QR}) in  $\frac{d}{dr}f(r)_{RN}=0$, we obtain the condition; $2-2r+3cr^{2}=0$, that has two real roots (horizons) denoted by $r_{+}$ and $r_{-}$, for $c<c_{crit}=1/6$ (where  $c_{crit}$ is obtained of (\ref{Cw})).

\begin{equation}
r_{\pm}=\frac{1\pm \sqrt{1-6c}}{3c}.
\end{equation}

When $r_{+}$ is considered in (\ref{QR}) ($q^{2}(r_{+},c)$), we obtain the same result reported in \cite{Malakolkalami:2015tsa} which is valid for $1/8 \leq c \leq c_{crit}$  and is given by ;

\begin{equation}\label{r+}
q^{2}_{r_{+}}=\frac{2}{27}\frac{-1+9c+\sqrt{-(6c-1)^{3}}}{c^{2}}.
\end{equation}

Now when $r_{-}$ is considered for $q^{2}(r_{-},c)$, we obtain a new condition that allows us to consider a larger range of values for $c$ between $0 < c \leq c_{crit}$, and $q^{2}(r_{-},c)$ is given by.

\begin{equation}\label{r-}
q^{2}_{r_{-}}=\frac{2}{27}\frac{-1+9c+\sqrt{(1-6c)^{3}}}{c^{2}}.
\end{equation}

The extremal of Bardeen BH-$\omega$ is obtained when we introduce (\ref{Gb}) in $\frac{d}{dr}f(r)_{Bar}=0$ , then the expression $4-3r+4cr^{2}=0$ is obtained and has two roots (horizons) $r_{1+}$ and $r_{1-}$ given by;

\begin{equation}\label{r+-}
r_{1\pm}=\frac{3 \pm \sqrt{9-64c}}{8c},
\end{equation} 

the roots are real for $c<c_{crit}=9/64$, where  $c_{crit}$ is obtained from (\ref{Cw1}), it is not possible to reproduce the result obtained in  \citep{Gha16} \citep{Ghaderi:2017yfr}, because it is only valid in ranges for $c$ where $g^{2}<0$. In the case of $r_{1+}$, $g^{2}(r_{1+},c)$ is valid in the range $1/8 \leq c \leq 9/64$. 

\begin{figure}[h]
\begin{center}
\includegraphics [width =0.45 \textwidth ]{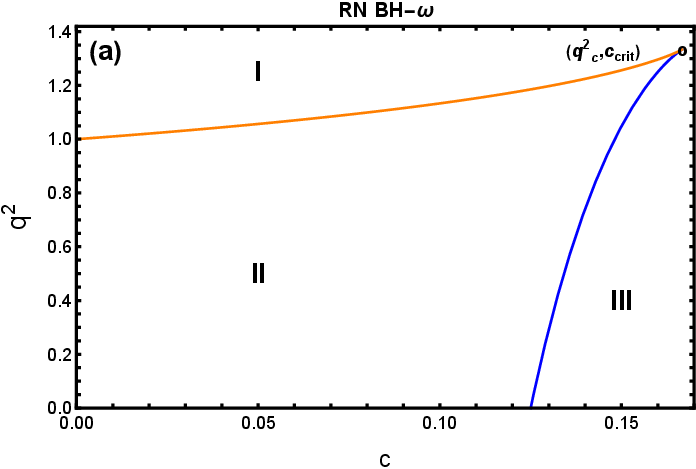}
\includegraphics [width =0.45 \textwidth ]{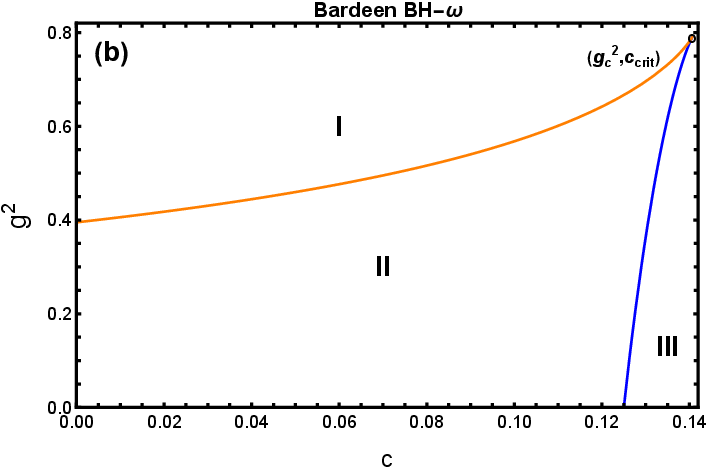}
\end{center}
\caption{ {\bf (a)} The behavior of $q^{2}$ as a function of $c$, is shown for $\omega=-2/3$. The regions I, II an III contain the values of $q^{2}$ and $c$ so that RN BH-$\omega$ has one horizon (I and III), three horizons (II) and in the boundaries, the black hole has two horizons. {\bf (b)} The behavior of $g^{2}$ as a function of $c$ is shown for $\omega=-2/3$. In the regions I and III Bardeen BH-$\omega$ has one horizon, in the region II it has three horizons and in the boundaries,  Bardeen BH-$\omega$ has two horizons.}\label{Fig2}
\end{figure}

In Fig. \ref{Fig2} {\bf (a)} the behavior of $q^{2}$ as a function of $c$ is shown for $\omega=-2/3$. In the regions I and III, the RN BH-$\omega$ has one horizon, while in the region II there are three horizons. The boundary of the regions I and II ($q^{2}(r_{-},c)$) as well as the boundary of regions II and III ($q^{2}(r_{+},c)$) represents the extremal RN BH-$\omega$. 

In Fig. \ref{Fig2} {\bf (b)} the behavior of  $g^{2}$ as a function of $c$ is shown for $\omega=-2/3$. In the regions I and III, the Bardeen BH-$\omega$ has one horizon, in the region II there are three horizons. In the boundary of regions I  and II ( $g^{2}(r_{1-},c)$) and in the boundary of  II and III ( $g^{2}(r_{1+},c)$), the Bardeen BH-$\omega$ has two horizons.

Depending on the values of the parameters $q^{2}$, $g^{2}$, and $c$, the number of the horizons may decrease from three to one, the cosmological (quintessence) horizon  $r_{q}$ never vanishes, when this unique horizon remains, the  BHs-$\omega$ describe a naked singularity.

\subsection{Black holes with $\omega=-1/2$}

Now we will focus on the particular case of $\omega= -1/2$. The Schw BH-$\omega$ has two horizons for $c<\sqrt{2/27}$. 

For $\omega= -1/2$ ,the extrema of RN BH-$\omega$ can be obtained from the condition $4-4r+5cr^{3/2}=0$, this can be solved analytically for $c \leq c_{crit}=8/(15 \sqrt{3})$, and  has two real positive roots for which $0< q^{2} \leq q^{2}_{c}= 9/5$. The two roots (horizons) are given by;

\begin{equation}
\widetilde{r}_{+}=\frac{2(64+8A^{1/3}+A^{2/3}-600c^{2})}{75A^{1/3}c^{2}},
\end{equation}

{\footnotesize
\begin{equation}
\widetilde{r}_{-}=\frac{1}{300c^{2}}\left(64+\frac{32(1-i\sqrt{3})(75c^{2}-8)}{A^{1/3}}-4(1+i\sqrt{3})A^{1/3}\right),
\end{equation}}

with {\footnotesize $A=512-7200c^{2}+16857c^{4}+375c^{3}\sqrt{3(675c^{2}-64)}$}. 

In the case of the Bardeen-BH-$\omega$ extreme, we obtain the equation $8-6r+7cr^{3/2}=0$ that can have two real positive roots for $c \leq c_{crit}=2/7$ and are given by;

 \begin{equation}
\widetilde{r}_{1+}=\frac{2(6B^{1/3}+(2B)^{2/3}+2^{4/3}(9-98c^{2}))}{49B^{1/3}c^{2}},
\end{equation}

\begin{figure}[h]
\begin{center}
\includegraphics [width =0.45 \textwidth ]{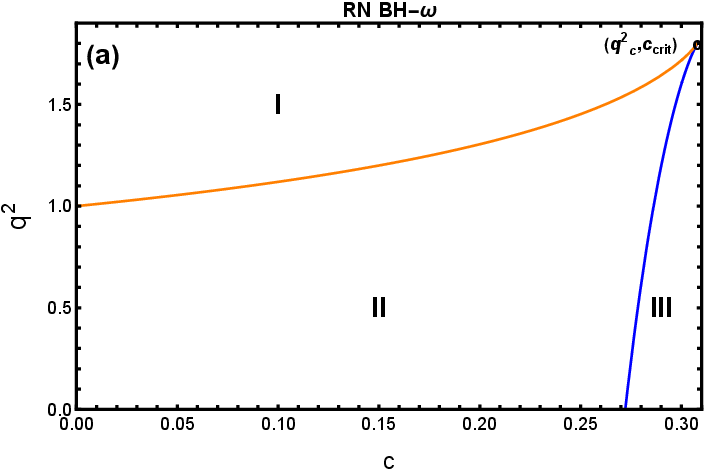}
\includegraphics [width =0.45 \textwidth ]{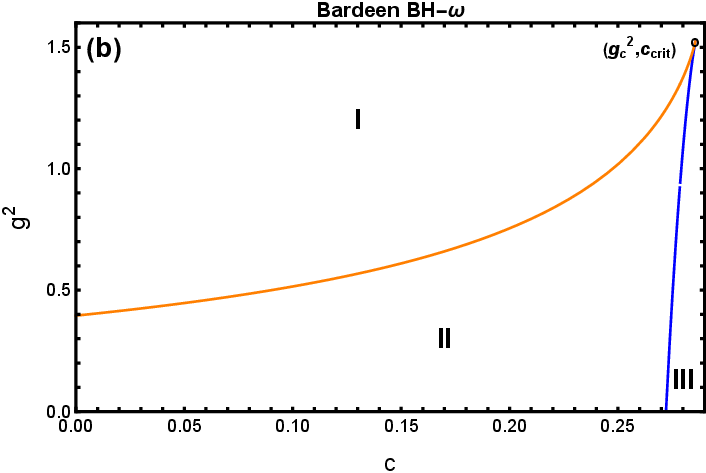}
\end{center}
\caption{{\bf (a)} The behavior of $q^{2}$ as a function of $c$ is shown for $\omega=-1/2$. The regions I, II an III contain the values of $q^{2}$ and $c$ so that RN BH-$\omega$ has one horizon (I and III), three horizons (II) and in the boundaries, the black hole has two horizons. {\bf (b)} The behavior of $g^{2}$ as a function of $c$ is shown for $\omega=-1/2$. The regions I and III Bardeen BH-$\omega$ has one horizon, in the region II has three horizons and in the boundaries, the Bardeen BH-$\omega$ has two horizons.}\label{Fig3}
\end{figure}

{\footnotesize
\begin{equation}
\widetilde{r}_{1-}=\frac{12B^{1/3}-(2B)^{2/3}(1+i\sqrt{3})-2^{4/3}(1-i\sqrt{3})(9-98c^{2})}{49B^{1/3}c^{2}},
\end{equation}}

where $B=54-882c^{2}+2401c^{4}+343c^{3}\sqrt{49c^{2}-4}$. In general the different roots satisfy the relation $0 < g^{2} \leq g^{2}_{c}= 32/21$, the behavior of $g^{2}$ as a function of $c$ is shown in Fig. \ref{Fig3} {\bf (b)}.

The behavior of $q^{2}$ as function of $c$ is shown in Fig. \ref{Fig3} {\bf (a)}. In the regions I and III, the RN BH-$\omega$ has one horizon, while in the region II it has three horizons. The boundary of regions I  and II ($q^{2}(\widetilde{r}_{-},c)_{RN}$) as well as the boundary of regions II and III ( $q^{2}(\widetilde{r}_{+},c)_{RN}$) represent the extremal case. 

The Fig. \ref{Fig3} {\bf (b)} shows regions I and III, where Bardeen BH-$\omega$ has one horizon, while in the region II there are three horizons. The boundary of regions I  and II ($g^{2}(\widetilde{r}_{1-},c)_{Bar}$) and the boundary of regions II and III ( $g^{2}(\widetilde{r}_{1+},c)_{Bar}$) represents the extremal Bardeen BH-$\omega$. 

Comparing the different regions, it is possible to conclude that $III_{\omega=-2/3} < III_{\omega=-1/2}$ and the range of $c$ for $\omega=-2/3$ is less than the range for $\omega=-1/2$.

In general the  horizons ($r_{in}$, $r_{out}$ and $r_{\omega}$) and extrema horizons ($r_{min}$ and $r_{max}$) satisfy the relation $r_{in} \leqslant r_{min} \leqslant r_{out} \leqslant r_{max} \leqslant r_{\omega}$ and there are three types of horizons (see for example \citep{Toshmatov:2015npp} ), in both cases $\omega=-2/3$ and $\omega=-1/2$, in short, they are:

\begin{itemize}

\item Type 1: The inner and outer horizons merge into a single horizon $r_{in}=r_{out}$, the boundary of regions I  and II.

\item Type 2: In this case, $r_{out}=r_{\omega}$ and $r_{in}< r_{-}$, the boundary of regions II  and III.

\item Type 3:  The three horizons merge into a single horizon, and the Kiselev black hole is known as a super-extremal black hole.

\end{itemize}

\section{Classical and semi-classical cross-sections}\label{3}

The scattering of light by  black holes is one way to study the geometry produced by these objects. Astronomers use the bending of light rays to measure the gravitational mass of clusters or galaxies. If a photon comes too close to the object, it will not escape, giving birth to a darker area known as the shadow of the black hole.

To investigate the scattering, it is necessary to find the scattering cross section, a  procedure used is the study of geodesics in the classical approximation. The geodesic analysis is  important because at very high frequencies the wave propagates along null geodesics \citep{Collins_1973}.

The test particles propagating along null geodesics are described by the Lagrangian density $ L=- \frac{1}{2}\dot{x}^{\mu}\dot{x}_{\mu}=0$, where "dot" denotes the derivative with respect to the affine parameter $\tau$. 

Using the form of the line element in (\ref{mfa}), the Lagrangian density is given by:

\begin{equation}\label{ec.la}
L=-\frac{1}{2}\left(-f(r)\dot t^2+\frac{\dot r^2}{f(r)}+r^2\dot\theta^2+r^2\sin^2\theta\dot\phi^2\right).
\end{equation}

We restrict the geodesic motion to the plane $\theta=\pi / 2$. The energy  $E=\frac{\partial L}{\partial \dot{t}}$ and the angular momentum $l=-\frac{\partial L}{\partial \dot{\phi}}$ of a test particle are conserved quantities, then the null geodesics are given by;

\begin{equation}
0=\frac{E^2}{f(r)}-\frac{\dot r^2}{f(r)}-\frac{l^2}{r^2}\,.
\end{equation}

Solving for $\dot{r}^2$, we obtain $\dot{r}^2=E^2-V_{eff}$, where $V_{eff}=f(r)\frac{l^2}{r^2}$. The presence of a maximum or a minimum in the effective potential $V_{eff}$ indicates that unstable or stable circular orbits are present. It is well known that null geodesics passing very close to an unstable circular orbit may be scattered through a deflection angle.

The circular orbits are determined from the conditions $V_{eff}^{'}(r_{c})=0$ and $ V_{eff}(r_{c})=E$ where $r_{c}$ is the radius of the circular orbit and the prime denotes derivative respect to $r$. Circular orbits satisfy the conditions;

\begin{equation}\label{Con1}
\frac{1}{\hat{b}^{2}}-\frac{f(r_{c})}{r_{c}^{2}}=0,
\end{equation}

and

\begin{equation}\label{Cond2}
2f(r_{c})-r_{c}f^{'}(r_{c})=0,
\end{equation}

where $\hat{b}=l/E$ is the impact parameter. The positive root of (\ref{Cond2}) corresponding to the radius of the unstable orbit (the turning point of radial motion) is substituted in  (\ref{Con1}). Then it is possible to obtain the impact parameter ($\hat{b}_{c}$) associated with the critical orbits. Null geodesic starting with the impact parameter $\hat{b}_{c}=\hat{b}$ end on the unstable orbit (rotates one or many time), when $\hat{b}<\hat{b}_{c}$ the null geodesic is absorbed by the BH and if $\hat{b}>\hat{b}_{c}$ the null geodesic is scattered by the BH. 

If we define $u=1/r$ and integrate $\dot{r}^2=E^2-V_{eff}$, we can obtain the expression;

\begin{equation}\label{Ang}
\left(\frac{du}{d\phi}\right)^{2}=\frac{1}{\hat{b}^{2}}-f(1/u)u^{2}.
\end{equation}

In the case of geodesics coming from the infinity to a turning point $u_{0}$ , the deflection angle is given by;

\begin{equation}\label{impact parameter}
\Theta(\hat{b})=2\phi(\hat{b})-\pi,
\end{equation}
where
\begin{equation}
 \phi(\hat{b})=\int_0^{u_0}du \left(\frac{1}{\hat{b}^2}-u^2 f(1/u)\right)^{-1/2}.
\end{equation}

The impact parameter $\hat{b}(\Theta)$, associated with the scattering cross-section is given by;

\begin{equation}\label{Sec}
 \frac{d\sigma}{d\Omega}=\frac{1}{\sin \Theta} \sum \hat{b}(\Theta)  \left| \frac{d\hat{b}(\Theta)}{d \Theta}\right|.
 \end{equation}
 
When we consider partial waves in the scattering phenomenon, it is necessary to consider the interference that occurs between partial waves with different angular momenta. This situation is not considered by the classical scattering (\ref{Sec}). The approximate method that considers the interference of the waves for large angles and high-frequency scalar plane waves ($\text{w} >> 1$) is the semi-classical approach (Glory scattering) \citep{Mat85}.

The unstable orbits give rise to a Glory in the backward direction. A Glory is a bright spot or halo that appears on--axis in the backward direction from the scatterer.
 
 The Glory approximation of the scalar scattering cross sections by spherically symmetric Black Holes is given by;

\begin{equation}
\frac{d\sigma_g}{d\Omega}=2\pi \text{w} \hat{b}_g^2 \left| \frac{d\hat{b}}{d\Theta}\right|_{\Theta=\pi}J_{2s}^2(\text{w} \hat{b}_g \sin\Theta),
\end{equation}

here, $\text{w}$ is the wave frequency. $J_{2s}^2$ stands for the Bessel function of first kind of order $2s$ where $s$ represents the spin, $s=0$ for scalar waves. The impact parameter of the reflected waves ($\theta \sim\pi$) is denoted by $\hat{b}_g$. As a semi-classical approximation, it is valid for $M\text{w} \gg 1$ ($M$ the mass of the BH)

\subsection{Classical and semi-classical scattering cross-sections with $\omega=-2/3$ and  $\omega=-1/2$ }

We calculate the classical and semi-classical scattering cross-sections for the black holes surrounded with quintessence considering the metrics (\ref{fs}), (\ref{fr}) and (\ref{fb}). Then, we analyze and compare the differences that may be similar but not identical. To carry out the analysis scattering cross-section, we consider that the three BHs-$\omega$ have two horizons (Type 1), for which the results of the previous section are taken.

Also, as the null geodesics movement can be located within the quintessence horizon, the impact parameter's numerical calculation ($\hat{b}_{c}$) is performed assuming this fact.

\begin{figure}[h]
\begin{center}
\includegraphics [width =0.45 \textwidth ]{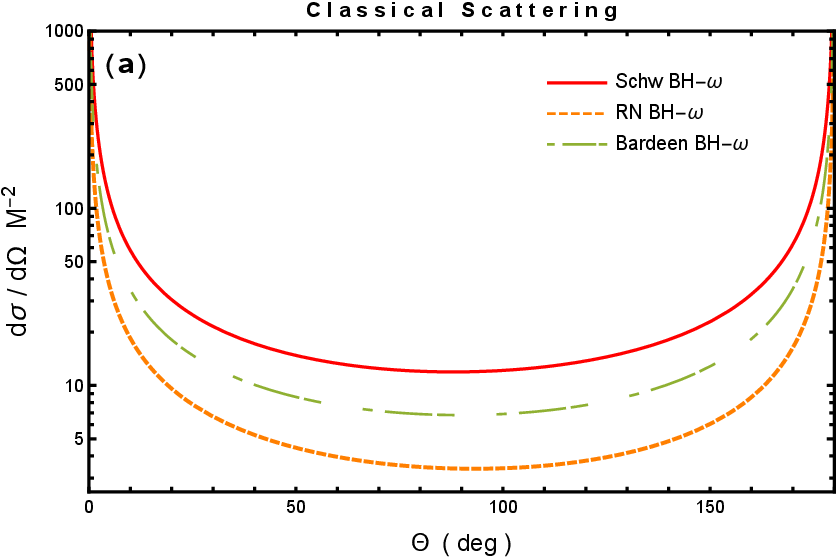}
\includegraphics [width =0.45 \textwidth ]{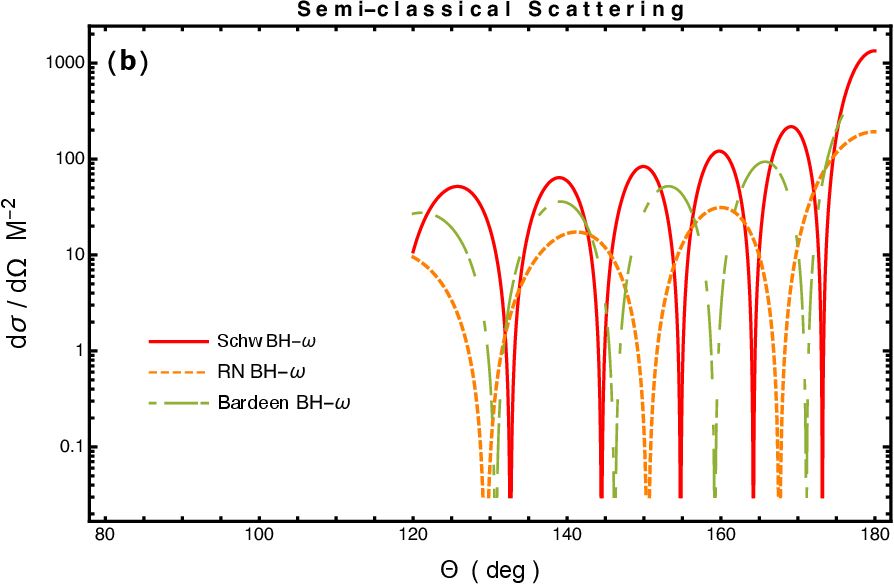}
\end{center}
\caption{{\bf (a)} Behavior of classical scattering cross-section for $\omega=-2/3$ with $c=0.1$ and $\text{w}M=2$. {\bf (b)} Behavior of semi-classical scattering cross-section for $\omega=-2/3$ with $c=0.1$ and $\text{w}M=2$.}\label{Fig4}
\end{figure}

The classical cross-sections for Schw BH-$\omega$ , RN BH-$\omega$, and Bardeen BH-$\omega$ are compared considering the case $\omega=-2/3$ in the Fig. \ref{Fig4} {\bf (a)}. In the  \ref{Fig4} {\bf (b)} semi-classical scattering cross-sections are compared. It is possible to note that there is no significant difference between the BHs-$\omega$ in the case of the classical scattering cross-section for small and large angles (see Fig. \ref{Fig4} {\bf (a)}).

Comparing the scattering cross-sections from Bardeen BH-$\omega$ and  RN BH-$\omega$, we observe that  the classical scattering cross-section for Bardeen BH-$\omega$ is greater than the one from RN BH-$\omega$ while in the case of  Schw BH-$\omega$, it is greater than Bardeen BH-$\omega$. These differences are compared by obtaining $ (\frac{d\sigma}{d\Omega})_{Schw}> (\frac{d\sigma}{d\Omega})_{Bardeen}>(\frac{d\sigma}{d\Omega})_{RN}$. These comparisons are obtained for both scattering cross-sections (classical and semi-classical) in the case $\omega=-2/3$.

\begin{figure}[h]
\begin{center}
\includegraphics [width =0.45 \textwidth ]{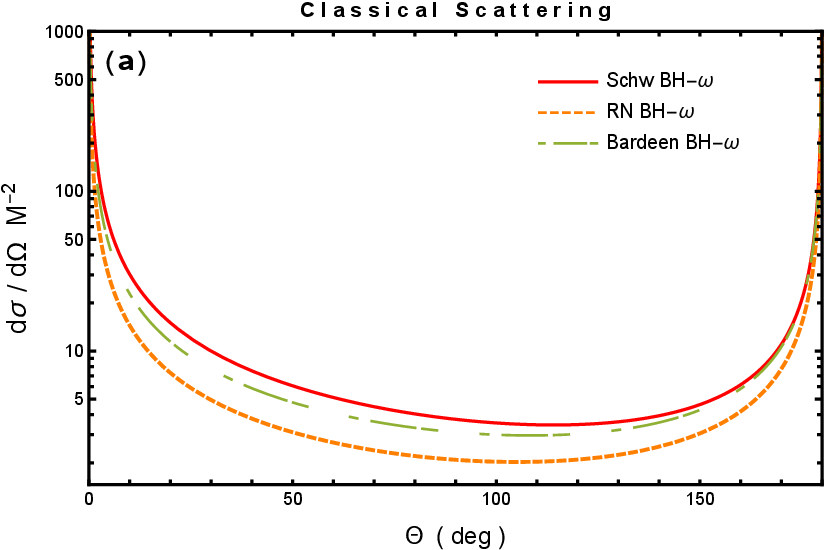}
\includegraphics [width =0.45 \textwidth ]{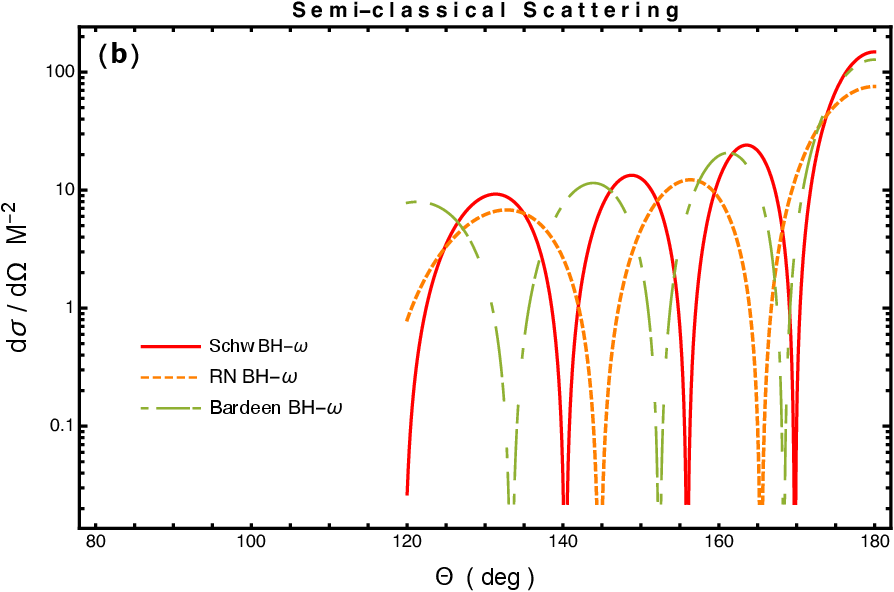}
\end{center}
\caption{{\bf (a)} Behavior of classical scattering cross-section for $\omega=-1/2$ with $c=0.1$ and $\text{w}M=2$. {\bf (b)} Behavior of semi-classical scattering cross-section for $\omega=-1/2$ with $c=0.1$ and $\text{w}M=2$.}\label{Fig5}
\end{figure}

When consider the case $\omega=-1/2$  the classical scattering cross-sections for Schw, RN, and Bardeen BHs-$\omega$ are shown in the Fig. \ref{Fig5} {\bf (a)}. The semi-classical scattering cross-sections are are shown in the Fig. \ref{Fig5} {\bf (b)}. We also consider only two horizons and the behavior is similar to the one in the case of $\omega =-2/3$. When we compare the Fig. \ref{Fig4} {\bf (a)} and Fig. \ref{Fig5} {\bf (a)}, where it is obtained $ (\frac{d\sigma}{d\Omega})_{(\omega=-2/3)}> (\frac{d\sigma}{d\Omega})_{(\omega=-1/2)}$, we see that the width of the interference fringes increases with $\omega=-1/2$ and in the case of $\omega=-2/3$ it decreases, and the difference between the BHs--$\omega$ is more noticeable with $\omega=-1/2$ (see Figs.  \ref{Fig4} {\bf (b)} and \ref{Fig5} {\bf (b)}).

\section{Absorption cross-section}\label{4}

An essential aspect of BHs is their accretion rate related to absorbing matter and fields. Accretion has an important role in the phenomenology of active galactic nuclei. Also, in the classical (high frequency) limit, absorption cross-sections are directly related to the shadows of BHs.

In the approximation for low frequencies, the absorption cross-section by spherically symmetric Black Holes equals the BH horizon area \cite{Higuchi:2001si}.

In the approximation for high frequencies, the absorption cross-section can be considered the classical capture cross-section of the null geodesics in massless scalar waves. In this limit the absorption cross-section is also called geometric cross-section $\sigma_{geo}=\pi \hat{b}_{c}^{2}$.

In \cite{Dec11} the author shows that in the case of high frequencies (eikonal limit), the oscillatory part of the absorption cross-section can be written in terms of the parameters of the unstable null circular orbits. 

Then  the absorption cross section  in the eikonal limit is defined by ;

\begin{equation}
\sigma_{osc}= - 4\pi \frac{\lambda \hat{b}_{c}^{2}}{\text{w}}e^{-\pi \lambda \hat{b}_{c}}\sin\left(\frac{2\pi \text{w},}{\Omega_{c}}\right)
\end{equation}

 where $\lambda$ is the Lyapunov \citep{Cardoso:2008bp} exponent given by;

\begin{equation}\label{Lyapunov}
\lambda^{2}=\frac{f(r_{c})}{2r_{c}^{2}}\left[2f(r_{c})-r_{c}^{2}f^{''}(r_{c})\right],
\end{equation}

$r_{c}$, the radius of the unstable null circular orbit and $\hat{b}_c$ is the critical impact parameter.

while the orbital angular velocity  is ;

\begin{equation}\label{angular1}
\Omega_{c}= \sqrt{\frac{f_c}{r_c^2}}.
\end{equation}

Then the absorption cross-section in the limit for high frequencies can be written as the sum of $\sigma_{osc}$ and $\sigma_{geo}$ (the sinc approximation)

\begin{equation}
\sigma_{abs} \approx \sigma_{geo} + \sigma_{osc}.
\end{equation}

\subsection{Absorption cross-section with $\omega=-2/3$ and  $\omega=-1/2$ }

\begin{figure}[h]
\begin{center}
\includegraphics [width =0.45 \textwidth ]{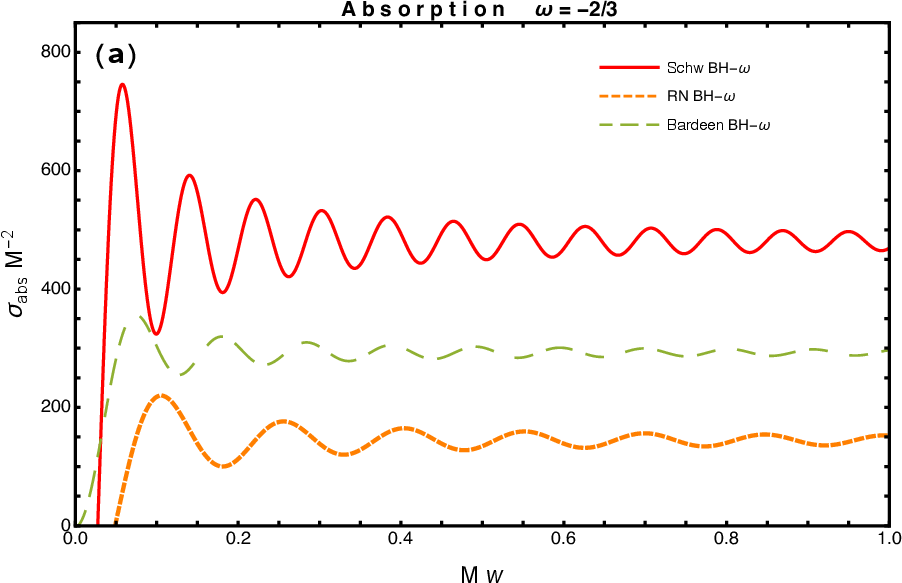}
\includegraphics [width =0.45 \textwidth ]{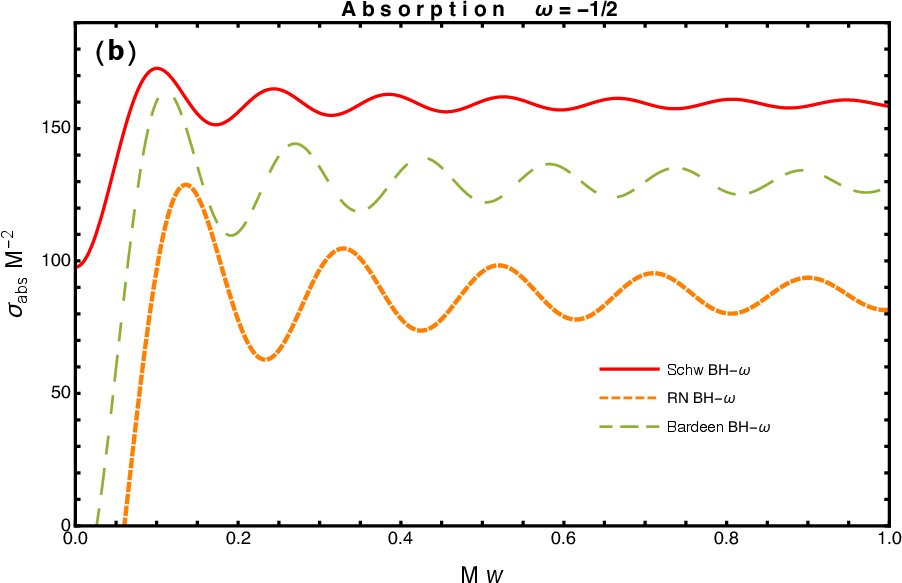}
\end{center}
\caption{{\bf (a)} Behavior of the Absorption cross-sections for $\omega=-2/3$ with $c=0.1$. {\bf (b)} Behavior of Absorption cross-sections for $\omega=-1/2$ with $c=0.1$}\label{Fig6}
\end{figure}

The absorption cross-sections for $\omega =-2/3$ are plotted in Fig. \ref{Fig6} {\bf (a)} and in the Fig. \ref{Fig6} {\bf (b)} for $\omega =-1/2$, both in the case of the sinc approximation. It is obvious that the amplitude of absorption cross-sections tend to a constant as $\text{w} M$ increases. Comparing the curves with the same value of $c=0.1$, we can observe that $\sigma_{abs}$ of Schw BH-$\omega$ is larger than the absorption cross-section of Bardeen and RN BHs-$\omega$ (as the case of the classical and semi-classical scattering cross-sections). It also shows that the difference between the curves is numerically significant with respect to their amplitude with larger values of $\text{w} M$ in both cases.

Moreover, for each value $\omega$, it is possible to mention that $\sigma_{abs_{-2/3}}>\sigma_{abs_{-1/2}}$, the corresponding absorption cross-section also starts from zero, reaches a maximum value $\sigma_{abs}$, and decreases asymptotically. 

\begin{figure}[h]
\begin{center}
\includegraphics [width =0.45 \textwidth ]{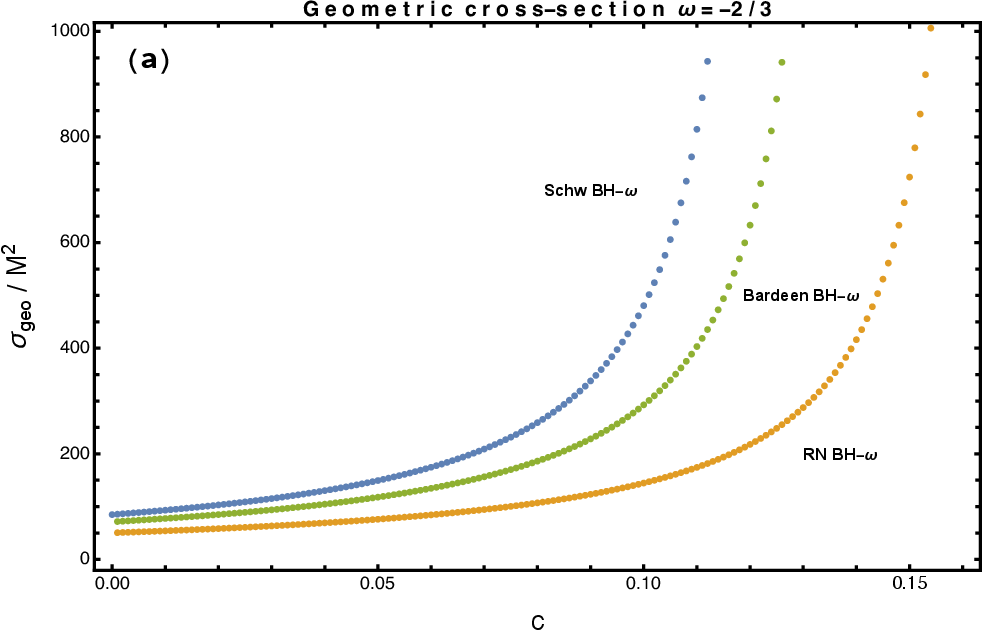}
\includegraphics [width =0.45 \textwidth ]{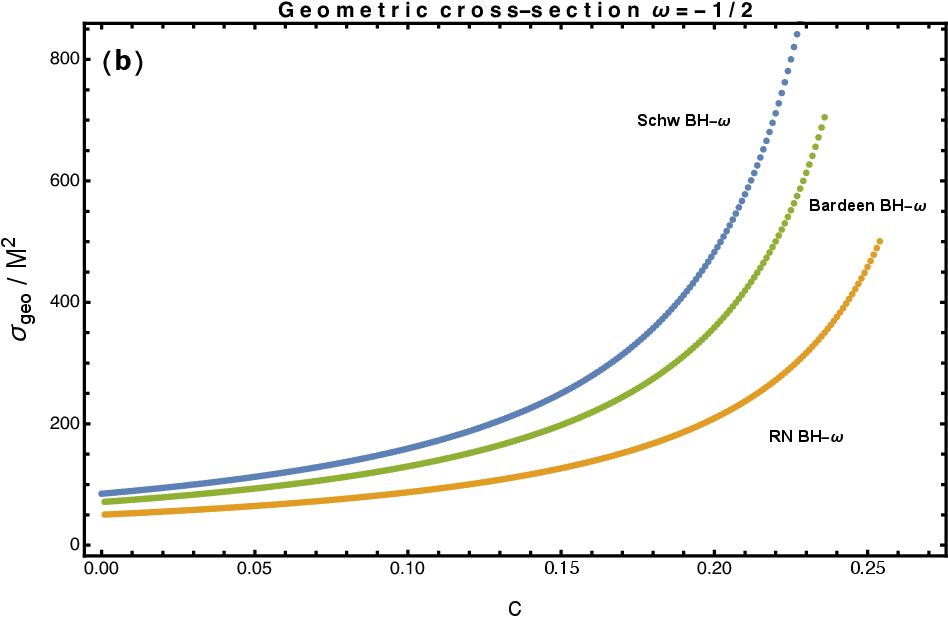}
\end{center}
\caption{{\bf (a)} Behavior of the Geometric  cross sections for $\omega=-2/3$. {\bf (b)} Behavior of Geometric  cross sections for  $\omega=-1/2$}\label{Fig7}
\end{figure}

Finally in the Fig. \ref{Fig7}, we compare the geometric cross-section $\sigma_{geo}$ of the different BHs-$\omega$, as a function of the parameter $c$. The case for $\omega=-2/3$ is shown in the Fig. \ref{Fig7} {\bf (a)} and for $\omega=-1/2$ in the Fig. \ref{Fig7} {\bf (b)}. In general the Schwarzchild BH-$\omega$ has a bigger geometric cross-section, compared with the Bardeen BH-$\omega$ and RN BH-$\omega$, in both cases $\omega=-2/3$ and $\omega=-1/2$, thus it is consistent with the behavior of the classical and semi-classical scattering cross-sections.

\section{Conclusions}\label{5}

In this paper, we study three regular black holes surrounded by quintessence in the special cases of $\omega=-2/3$ and $\omega=-1/2$, we have also focused our attention in these cases that enable a relatively simple treatment of the properties of the black holes surrounded by quintessence. The three black holes considered are the Schwarzschild, Reissner-Nordstr\"{o}m and Bardeen black holes. We also present the critical values of $c_{c}$ (normalization factor) and charge ($q$ and $g$) parameters in terms of $\omega$. We obtain that in the presence of the quintessence field, a higher charge black hole is formed in the case of RN BH-$\omega$ and Bardeen BH-$\omega$.

The classical and semi-classical scattering cross-sections are compared for the three black holes in the case where the black holes have two horizons and are of type 1. We observe that the classical scattering cross-section for Bardeen BH-$\omega$ is greater than RN BH-$\omega$ while in the case of the Schw BH-$\omega$, it is more significant than Bardeen BH-$\omega$.

For the absorption cross-section, the absorption cross-section for Schw BH-$\omega$ is larger than the absorption cross-section of Bardeen and RN BHs-$\omega$ and it decreases asymptotically with large values of $\text{w}M$.

A comment that the classical scattering cross-section deserves, is that when the scattering angle value increases the cross-section has a minimum value and then increases for larger angles.

Regarding the Geometric cross-sections, it is possible to mention that when the normalization factor $c$ increases, the geometric cross-sections also increase and the highest value always corresponds to the Schwarzschild BH-$\omega$.

The Figs. \ref{Fig4} {\bf (a), (b)} and \ref{Fig5} {\bf (a), (b)} show that the classical and semiclassical scattering cross-sections are greater for $\omega=-2/3$. A similar behavior is observed for the absorption section (Fig. \ref{Fig6}). So, we expect that the sections are more significant in the vicinity of $\omega\to -1$. 

Finally, we can mention that this type of study is of utmost importance since dark matter does not interact with photons, and it would not deflect light directly. It is also difficult to think that dark energy could affect the surroundings of a black hole. But the metric's modification affects the path of light rays. So then, for example, we can observe the modification of the shadow of a black hole by dark matter.

\acknowledgments

The authors acknowledge the financial support from PROMEP project UAEH-CA-108 and  SNI-CONACYT, M\'exico.

\end{document}